\documentclass[seceq,jjapprn]{jjap}
\usepackage{graphicx}
\usepackage{color}

\title{Sudden Suppression of Electron-Transmission Peaks in Finite-Biased Nanowires}
\author{Shigeru~\textsc{Tsukamoto}$^1$, Masakazu~\textsc{Aono}$^{1,2}$ and Kikuji~\textsc{Hirose}$^2$}
\inst{$^1$The Institute of Physical and Chemical Research, 2-1 Hirosawa, Wako, Saitama 351-0198, Japan\\
$^2$Department of Precision Science and Technology, Osaka University, Suita, Osaka 565-0871, Japan}
\abst{Negative differential conductance (NDC) is expected to be an essential property to realize fast switching in future electronic devices. We here present a thorough analysis on electron transportability of a simple atomic-scale model consisting of square prisms, and clarify the detailed mechanism of the occurrence of NDC phenomenon in terms of the changes of local density of states upon applying bias voltages to the electrodes. Boosting up bias voltages, we observe sudden suppression of transmission peaks which results in NDC behavior in the current-voltage characteristic. This suppression is explained by the fact that when the bias voltage exceeds a certain threshold, the conduction channels contributing to the current flow are suddenly closed up to deny the electron transportation.}
\kword{quantized conducntance, negative differential conductance, electron transmission, local density of states, finite-biased nanowire, Lippmann-Schwinger equation}
\begin{document}
\maketitle
\sloppy

\section{Introduction}
As the downsizing of electronic devices progresses more and more, it becomes more important to understand electron-transport properties of atomic-scale structures where the quantum effect predominates. Several unique properties of electron transport in such minute structures have been demonstrated by many experimental and theoretical studies\cite{JMRuitenbeek00,JMKrans95,JChen99,SDatta95,NDLang00,HSSim01,NKobayashi01,STsukamoto01,IWLyo89,NDLang97,MDVentra01,STsukamoto02}. In particular, negative differential conductance (NDC), which is an extremely unique electron-transport phenomenon, attracts our attentions\cite{IWLyo89,NDLang97,MDVentra01,STsukamoto02}, because NDC is expected to be an essential property that allows fast switching in a certain type of future electronic devices.

The evidence of NDC was first observed in minute structures ($\sim$1 nm) by Lyo and Avouris using the scanning tunneling microscopy (STM) and spectroscopy\cite{IWLyo89}. They measured the current-voltage (I-V) characteristics of the system involving a tunnel diode configuration formed by a specific site on boron-exposed Si(111) surface and an STM tip, and found NDC phenomenon occurring at a certain minority site, e.g., a defect site missing the surface Si atom. Theoretically, Lang explored NDC behavior in the I-V characteristics of an atomic contact consisting of Al-Br pairs attached to flat metal surfaces, and predicted NDC in a distance between the Al atoms of 8.6 a.u., a distance larger than twice the covalent radius of the Al atom\cite{NDLang97}. Moreover, Ventra {\it et al.}~developed a new fact that NDC phenomenon occurs in the benzene-ring family of molecules, taking the effect of temperature on geometrical and electronic structures into account\cite{MDVentra01}.

Recently, Tsukamoto and Hirose demonstrated on the basis of the first-principles theoretical study that NDC emerges in a single-row Na nanowire sandwiched by a couple of Na atomic plinths on flat metal surfaces, and that this unique NDC is attributed to the finding that some electron-transmission peaks are suddenly suppressed when the applied bias voltage exceeds a certain threshold value\cite{STsukamoto02}. In addition, they discussed in Ref.~12 the mechanism of this sudden suppression in terms of the changes of the effective potential along the nanowire (voltage drop) and of the conduction channels, as follows: As the bias voltage is boosted up, intensive voltage drop is formed between the nanowire and the negatively biased plinth, and then the onset energy of the electron transmission through the plinth, where the electron-conduction channel begins to open, moves to higher energies, and electrons with the low energy which are allowed to go through the plinth for bias voltages lower than the threshold come to be refused to transmit through the plinth. Consequently, the transmission peaks observed at the low energies are drastically suppressed at the bias voltage exceeding the threshold.

In this paper, we present a thorough analysis of the mechanism how the sudden suppression of electron-transmission peaks as an origin of NDC takes place. Employing a simple model involving two rectangular-parallelepiped prisms with different thickness, we examine electron-transport properties of the model by the method making use of the Lippmann-Schwinger equation proposed by Lang\cite{NDLang95}. A number of important results are derived from the present analysis: First, the sudden suppression of the transmission peaks and the resultant NDC behavior in the I-V characteristic are observed even in the present simplified model. Hence, one can conclude that NDC is a {\it universal} and {\it material-independent} phenomenon. Second, the sudden suppression of transmission peaks is brought about by the fact that the conduction channels contributing to the current flow at low bias voltages are suddenly closed up when the bias voltage exceeds a certain threshold value. Finally, the cutting off of the conduction channels is explained by the shift of the local density of states (LDOS) to higher energies with the rising of the bias voltage.

The paper is organized as follows: In Sec.~\ref{sec:Model} we describe a simplified model employed here and calculation conditions used for evaluation. The relation between the sudden suppression of the transmission peaks and the NDC behavior in I-V characteristic is demonstrated in Sec.~\ref{sec:Suppress}. The reason why this sudden suppression takes place is discussed in Sec.~\ref{sec:LDOS} on the basis of the local electron transportability. In Sec.~\ref{sec:New} we make mention of the spatial distributions of the LDOS including those of newly generated states due to the combination of the two prisms. Section \ref{sec:Conclusion} is devoted to concluding remarks.

\section{Calculation Model}
\label{sec:Model}
In Ref.~12 was stressed the argument that what plays an important role in the occurrence of NDC phenomenon is not only the single-row atomic nanowire but also the negatively biased atomic plinth attached to one end of the nanowire. In order to explore the universality of NDC mechanism, it is therefore necessary to adopt the model which well describes characteristics of these two indispensable components and is so simplified as to analyze readily electron-transport properties. Then we employ the model composed of two rectangular-parallelepiped prisms with different widths as well as a couple of semi-infinite jellium electrodes (Fig.~\ref{fig:Figure1}). The wider of the two prisms mimics the negatively biased plinth inserted between the left electrode and the single-row atomic nanowire, and the narrower prism the nanowire. The semi-infinite jellium electrodes are included as the electron reservoirs sandwiching this combined prism, thereby we can evaluate current flows and electron transport properties, and can apply arbitrary bias voltages to the system through assigning effective potentials to the respective prism parts independently.

It was further found in Ref.~11 that the intensive voltage drop, which is essential for NDC phenomenon, appears at the interface region between the single-row atomic nanowire and the negatively biased plinth,\cite{MBrandbyge99} and that each of the electric potentials within the negatively and positively biased sides of the interface region is almost constant. According to these findings, the intensive voltage drop is assumed to occur at the interface between the two prisms, and the effective potential in the left (right) side of the interface, i.e., the left electrode and wide prism part (the right electrode and narrow prism part), is taken to be constant. Thus we expect that the electron-transmission peaks are suddenly suppressed upon boosting up the bias voltage, if NDC is a universal phenomenon independent of materials.

In the numerical calculations, the widths of the wide and narrow prisms, $W_1$ and $W_2$, are taken to be 10.0 a.u.~and 6.0 a.u., respectively, and the lengths of the respective prisms, $L_1$ and $L_2$, to be 20.0 a.u. The simplified model is placed in the supercell on which periodic boundary conditions are imposed in the directions perpendicular to the combined prism ($x$ and $y$ directions) and nonperiodic boundary condition in the direction parallel to it ($z$ direction). Therefore, wave functions are expressed by the Laue representation. The side length of the supercell in the $x$ and $y$ directions is chosen to be 24.0 a.u.~and the number of the plane waves in these directions to be 24$^2$, which corresponds to a cutoff energy of 7.4 Ry. We use only the $\Gamma$ point in the two-dimensional Brillouin zone because of a large side length of the supercell adopted here. The grid size in the $z$ direction is taken to be 0.5 a.u.

The effective potential in the right side of the interface is defined as zero with/without the application of the bias voltage, and the potential in the vacuum region is fixed at 13.6 eV. For absence of the bias voltage, the effective potential in the left side of the interface is set to be equal to that in the right side, and the Fermi levels of the left and right electrodes are assigned to be 4.65 eV. For the presence of the finite bias voltage, both the effective potential and the Fermi level of the left side are raised up according to the applied bias voltage.

\section{Suppression of Transmission Peaks and the Occurrence of NDC Phenomenon}
\label{sec:Suppress}
It was reported in Ref.~12 that the sudden suppression of the electron-transmission peaks is recognized when the bias voltage exceeds a certain threshold value where the NDC phenomenon emerges. We first investigate whether even in the present simplified model the sudden suppression of the transmission peaks is also observed or not. Figure~\ref{fig:Figure2} indicates the electron transmissions through the combined prism at the bias voltages of 0.0, 1.20, 2.56, 3.10 and 3.65 V. The transmission curves are drawn as a function of the incident electron energy measured from the effective potential of the right electrode. It is obviously seen that the transmission peak clearly observed at the energy of 4.65 eV for the bias voltage less than 2.56 V is suddenly suppressed for the bias voltage of 3.10 V (A$\rightarrow$A'). Further, one more peak standing at 5.20 eV for the bias voltage of 2.56 V is also suppressed when the bias voltage is raised up to 3.65 V (B$\rightarrow$B').

We next examine the I-V characteristic of the combined prism around the bias voltage where the sudden suppression of the transmission peaks emerges. The shaded area in each panel of Fig.~\ref{fig:Figure2}, which is partitioned off by the Fermi levels of the both electrodes, corresponds to the voltage window where incident electrons contribute to the current. The current flow is evaluated by the integration of the transmission over the energy in the voltage window. Figure \ref{fig:Figure3} shows the I-V characteristic (solid line) of the combined prism and its differential conductance (broken line). The differential conductance is defined as the derivative of the current with respect to the bias voltage. In Fig.~\ref{fig:Figure3}, the following points are remarkable: One is that a rapid falling down of the differential conductance is observed for the bias voltages from 2.5 V to 3.1 V in which the first sudden suppression of the transmission peak (A$\rightarrow$A') occurs. The other is that the differential conductance becomes negative as the bias voltage is further boosted up to $\sim$3.7 V, where the current flow decreases against the increasing of the bias voltage. Since the transmission peak B just suppresses at the bias voltage of 3.65 V, as shown by B' in Fig.~\ref{fig:Figure2}, one evidently recognizes that the NDC phenomenon is induced by the sudden suppression of the electron-transmission peaks within the voltage window.

A series of the unusual behavior in the I-V characteristic of the combined prism can be understood by the increase/decrease of the integrated electron transmission, namely the increasing of the integration range (the voltage window) with the rising of the bias voltage, and the decreasing of the integrand (the electron transmission) originating from the sudden suppression of the transmission peaks. In this way, the differential conductance is determined by the competition between the increasing of the voltage window and the decreasing of the electron transmission.

\section{LDOS Analysis on the Sudden Suppression of Transmission Peaks}
\label{sec:LDOS}
We here examine the mechanism of the sudden suppression of the electron-transmission peaks which originates NDC behavior, based on the analysis of the LDOS of the combined prism. Figure \ref{fig:Figure4} shows the LDOS of electrons incident from the left electrode for the bias voltages of 0.0, 1.20, 2.56, 3.10 and 3.65 V. Solid (broken) curves represent the LDOS inside the narrow (wide) prism part. The LDOS curves are drawn as a function of the incident electron energy measured from the effective potential of the right electrode. As the bias voltage is boosted up, the LDOS curve in the wide prism part shifts to higher energies, because the effective potential inside the left electrode and wide prism part is raised up in proportion to the applied bias voltage. On the other hand, the effective potential of the right electrode and narrow prism part remains fixed for the rising of the bias voltage, and therefore no shift of the LDOS is to be brought about, i.e., the leftmost LDOS peak of the narrow prism part should appear at the energy of 4.65 eV for any bias voltage. However, no one can observe the leftmost peak at 4.65 eV for the bias voltages of 3.10 and 3.65 V. For these bias voltages, the leftmost peak of the LDOS in the narrow prism part looks as if it moves to higher energies, similarly to the LDOS in the wide prism part. Actually, the peak A of the LDOS inside the narrow prism part does not shift to the higher energies but becomes invisible as indicated by A$\rightarrow$A' in Fig.~\ref{fig:Figure4}, due to the blockade of electrons incident from the left electrode (this blockade can be clearly observed in the spatial LDOS distributions which will be shown later in Fig.~\ref{fig:Figure6}). One more vanishing of the LDOS peak at the energy of 5.20 eV, denoted by B$\rightarrow$B' in Fig.~\ref{fig:Figure4}, is also found with the boosting of  bias voltage up to 3.65 V. This vanishing of the respective LDOS peaks is in accordance with the sudden suppression of the corresponding transmission peaks shown in Fig.~\ref{fig:Figure2}. 

Let us enter into a discussion of details about the mechanism of the sudden suppression of the transmission peaks. We first remark that the electron transmission of the whole combined prism is understood from the {\it local} electron transportability of each prism part, which has a close relation to the LDOS of each prism part. It is common knowledge that the first channel of electron transmission becomes open at the energy of the first resonance state. So, we can easily see from Fig.~\ref{fig:Figure4} that the first local transmission channel of the {\it wide} prism part opens at the energy of 2.05, 3.20, 4.65, 5.05 and 5.65 eV for bias voltages of 0.0, 1.20, 2.56, 3.10, and 3.65 V, respectively, where the leftmost LDOS peak of the wide prism part stands for the first resonance state. On the other hand, the first local transmission channel of the {\it narrow} prism part is to open at 4.65 eV for all bias voltages because of the fixed effective potential inside the narrow prism part. (In the case of the bias voltage of 2.56 V in Fig.~\ref{fig:Figure4}, the onset of the local transmission of the narrow prism part comes about at the energy slightly lower than 4.65 eV due to a newly generated state S, which will be discussed in Sec.~\ref{sec:New}.)

We next show in Fig.~\ref{fig:Figure5} the illustrations of the local electron transportability of the two prism parts and electrodes arranged along the $z$ direction for the bias voltages of 0.00, 1.20, and 3.10 V. The thick line represents the first resonance state localized in each region, and the shaded area the energy range that the local transmission channel is open. In the following, we exhibit how electrons incident from the left electrode pass through the whole combined prism for the respective bias voltages:

(i) In the case that no bias voltage is applied (Fig.~\ref{fig:Figure5}(a)), the onset energy of the local transmission of the wide prism part is lower than 4.65 eV of the narrow one. Only electrons with the energy more than 4.65 eV can pass through both of the prism parts, since the electron transmission of the whole combined prism is controlled by a prism part with lower local transportability (higher onset energy) of the two, and resultant electron transmission of the whole system has the first onset at the energy of 4.65 eV, as seen in the top panel (the case of V$_B$=0.00V) of Fig.~\ref{fig:Figure2}. 

(ii) The case of the bias voltage of 1.20 V, a typical example of the finite bias voltage less than 2.56 V, is presented in Fig.~\ref{fig:Figure5}(b). The energy of the first resonance state of the wide prism part lifted up in proportion to the applied bias voltage, 3.20 eV, still remains below that of the narrow one, 4.65 eV, and therefore the narrow prism part keeps the initiative in the resultant electron transmission. The onset of the electron transmission of the whole combined prism thus keeps staying at 4.65 eV for the increasing of the bias voltage.

(iii) As the bias voltage is further boosted up to 3.10 V, the first resonance state in the wide prism part overtakes that of the narrow one at last to take place at the energy of 5.05 eV (Fig.~\ref{fig:Figure5}(c)). Then, the initiative in the electron transmission of the whole combined prism is passed to the wide prism part from the narrow one. Electrons with the energy of 4.65 eV incident from the left electrode are reflected at the wide prism part, i.e., the blockade of the incident electrons is induced to bring about the vanishing of the corresponding LDOS peak in the narrow prism part marked by A$\rightarrow$A' in Fig.~\ref{fig:Figure4}. Consequently, the electron-transmission peak at the energy of 4.65 eV suddenly suppresses at the bias voltage of 3.10 V, as indicated by A$\rightarrow$A' in Fig.~\ref{fig:Figure2}.

According to the same argument as the suppression of A$\rightarrow$A', the other noticeable suppression at the energy of 5.20 eV, B$\rightarrow$B', is also explained by the local transportability: Incident electrons with the energy of 5.20 eV is blocked out by the wide prism part, because the onset of the local transmission of the wide prism part appears at the energy of 5.65 eV for the bias voltage of 3.65 V (see Fig.~\ref{fig:Figure4}), and therefore the transmission peak at the energy of 5.20 eV suddenly suppresses. Thus the mechanism of the sudden suppression of the electron-transmission peaks is clarified by this consideration of the local electron transportability.

\section{Spatial LDOS Distributions and Newly Generated States}
\label{sec:New}
Figure \ref{fig:Figure6} shows the contour plots of the LDOS of electrons incident from the left electrode for the bias voltages of 0.0, 1.20, 2.56, 3.10, and 3.65 V. From the LDOS distributions labeled A at the electron energy of 4.65 eV, the blockade of electrons incident from the left electrode is found to occur near the entrance of the wide prism part upon boosting the bias voltage up to 3.10 V; the absence of electrons in the narrow prism part due to the blockade is clearly recognized in the cases of 3.10 and 3.65 V, while for the bias voltage of 2.56 V and below, electrons are present in the narrow prism part and reach the right electrode to contribute to the current flow. A series of the LDOS distributions at the electron energy of 5.20 eV labeled B also demonstrate that the blockade of electrons is not observed until the bias voltage is boosted up to 3.65 V. As regards to the LDOS distributions labeled C, the blockade of the incident electrons with the energy of 5.65 eV is not recognized at the bias voltage of 3.65 V yet. One can see that the LDOS distributions labeled C correspond to the second resonance states with a node at the center in the narrow prism part for all bias voltages. The independence of the LDOS distribution in the narrow prism part on the bias voltage is the definite evidence that no shift of the LDOS is brought about in the narrow prism part, which was already mentioned in Sec~\ref{sec:LDOS}.

The electron transmission for the bias voltage of 2.56 V shown in Fig.~\ref{fig:Figure2} exhibits the somewhat different behavior from the other curves, i.e., one can see newly generated transmission peaks S and B at the energies of 4.50 and 5.20 eV, respectively. Here we will confine our attention to the LDOS (Fig.~\ref{fig:Figure4}) and the spatial distributions of the LDOS (Fig.~\ref{fig:Figure6}) regarding these additional transmission peaks. At the bias voltage of 2.56 V in Fig.~\ref{fig:Figure4}, the resonance state at the energy of 4.65 eV of the wide prism part collides with that of the narrow prism part to generate two new states delocalized over both of the two prisms; one is at the energy of 4.50eV, slightly lower energy than 4.65 eV, and the other at the 5.20 eV. The spatial distributions of these additional states of electrons incident from the left electrode are depicted in the panels labeled S and B in Fig.~\ref{fig:Figure6}(c). One can see that there is an antinode at the interface of the two prism parts in each of these states in Fig.~\ref{fig:Figure6}(c), while a node at the interface in the other panels A and C, and therefore the additional states S and B with an antinode are not generated until the resonance states in the two prism parts hybridize with each other.

\section{Conclusion}
\label{sec:Conclusion}
In conclusion, in order to clarify the detailed mechanism of the occurrence of NDC and illustrate its universality independent of materials, we presented a thorough theoretial analysis of electron-transport properties of the combined prism using the Lippmann-Schwinger equation. Our calculation analyses provided us with some interesting results: Even the simple model composed of the two rectangular-parallelepiped prisms with different thickness exhibits the sudden suppression of the electron-transmission peaks, and the resultant NDC behavior in the I-V characteristic is induced by this sudden suppression. The increasing of the bias voltage brings about the shift of the onset energy of the local transmission of the wide prism part according to the rising of the electric potential of the left electrode. For low bias voltages, the narrow prism part predominates in the electron transmission of the whole system, because the onset energy of the local transmission of the narrow prism part is higher than that of the wide prism part. As the bias voltage exceeds a certain threshold value where the local-transmission onset of the wide prism part coincides with that of the narrow prism part, the wide prism part becomes predominant in the electron transmission. When the bias voltage is boosted up further, incident electrons that can reach the opposite electrode for the lower bias voltage come to be blocked in the wide prism part. Thus, the sudden suppression of the electron transmission emerges even in the present simplified model with the intensive voltage drop at the interface between the two prism parts, and this fact gives the evidence for the universality of NDC phenomenon.

\acknowledgements{%
Numerical calculations were performed using the SX-5 at Cyber Media Center, Osaka University. This work was supported by a Grand-in-Aid for COE Research (No.~08CE2004) from the Ministry of Education, Culture, Sports, Science and Technology.
}


\begin{halffigure}
\caption{Schematic representation for the combined prism connected to a couple of semi-infinite jellium electrodes. The effective potential inside the wide (narrow) prism part is set to be equal to that inside the left (right) electrode.}
\label{fig:Figure1}
\end{halffigure}

\begin{halffigure}
\caption{Transmissions of the combined prism as a function of the incident electron energy measured from the effective potential inside the right electrode. The shaded areas show the voltage windows. $E_F^{L(R)}$ represents the Fermi level of the left (right) electrode. The curve in each panel corresponds to the bias voltage indicated within.}
\label{fig:Figure2}
\end{halffigure}

\begin{halffigure}
\caption{Current-voltage characteristic of the combined prism (solid line) and its differential conductance (broken line).}
\label{fig:Figure3}
\end{halffigure}

\begin{halffigure}
\caption{LDOS of electrons incident from the left electrode inside each prism part. The incident electron energy is measured from the effective potential of the right electrode. Solid (broken) lines represent the LDOS of the narrow (wide) prism part. The curves in each panel correspond to the bias voltage indicated within.}
\label{fig:Figure4}
\end{halffigure}

\begin{halffigure}
\caption{Schematic representations for the mechanism of the sudden suppression of the electron transmission peaks. Three series of the local electron transportability along the z axis for the bias voltages of 0.00, 1.20, and 3.10 V are shown in (a), (b) and (c), respectively. The shaded areas are the energy ranges where the local transmission channels open. The thick lines represent the first resonance states of the respective regions.}
\label{fig:Figure5}
\end{halffigure}

\begin{fullfigure}
\caption{Spatial distributions of the LDOS of electrons incident from the left electrode. The panels for the bias voltage of 0.00, 1.20, 2.56, 3.10, and 3.65 V are arranged from top to bottom. The incident electron energies are 4.50, 4.65, 5.20, and 5.65 eV for the states S, A, B, and C, respectively. The contour lines are drawn in the logarithmic scale. The thick lines represent the outline of the model we employed.}
\label{fig:Figure6}
\end{fullfigure}

\makefigurecaptions
\begin{center}
\includegraphics{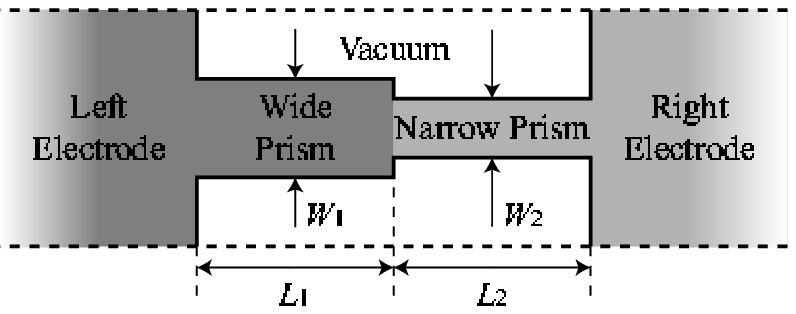}
\clearpage
\includegraphics{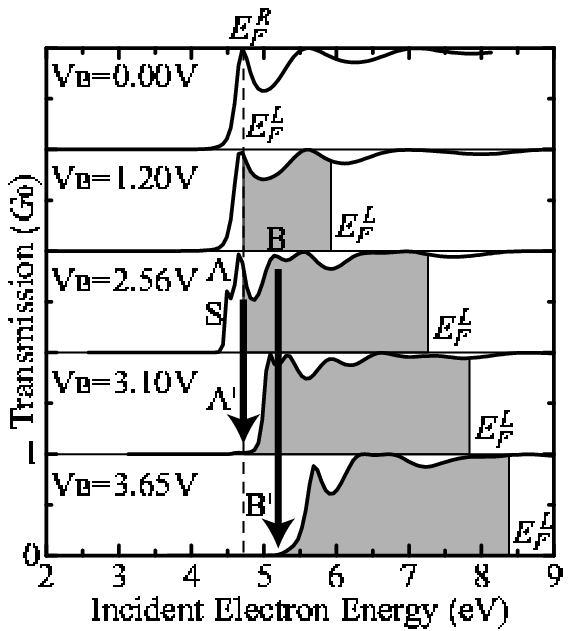}
\clearpage
\includegraphics{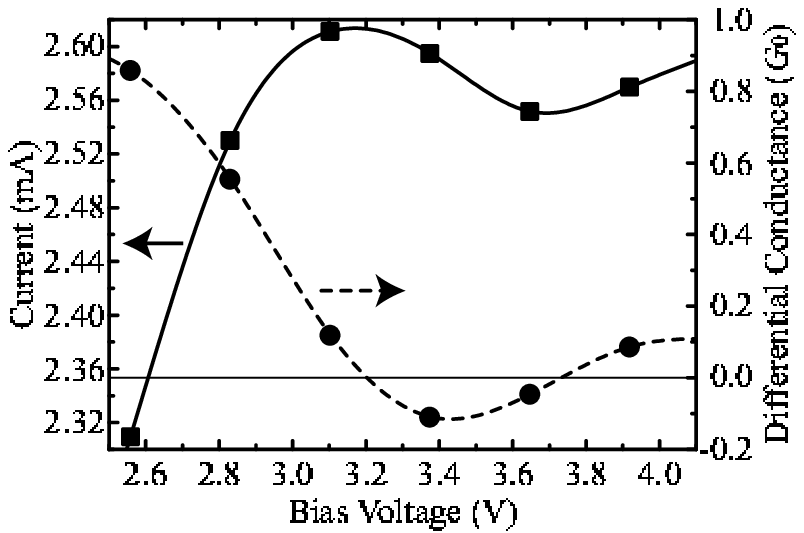}
\clearpage
\includegraphics{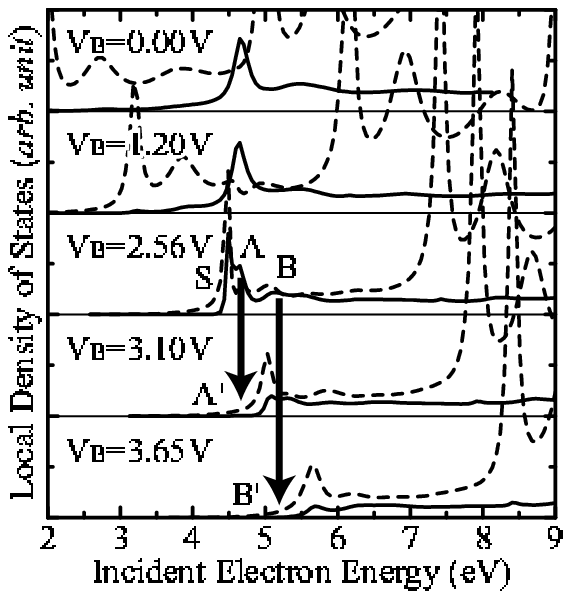}
\clearpage
\includegraphics{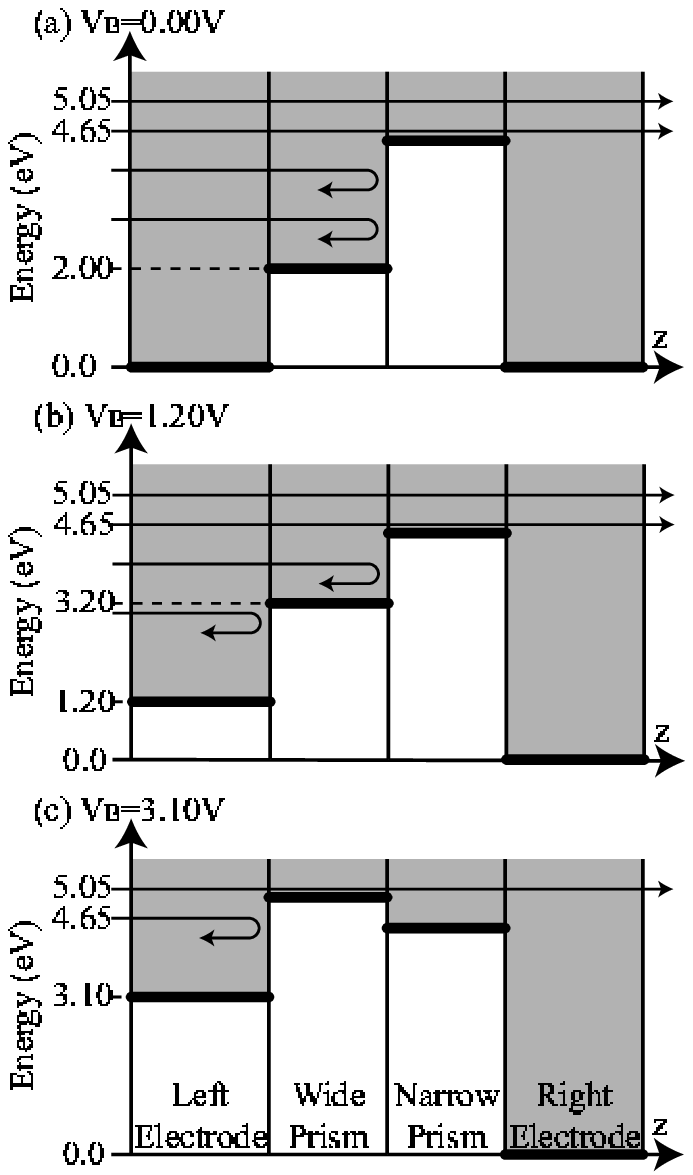}
\clearpage
\includegraphics{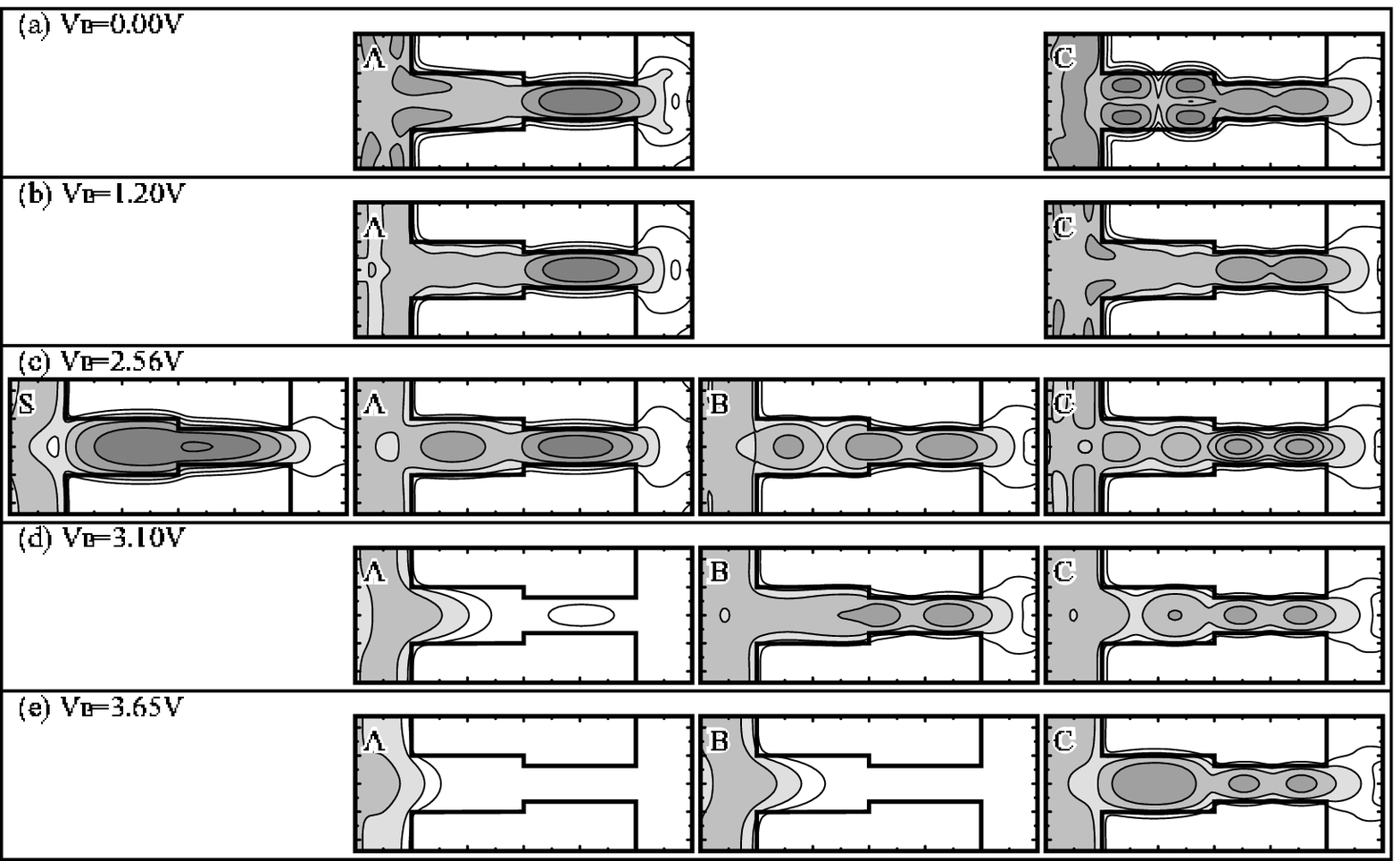}
\end{center}
\end{document}